# Computerized Tomography Pulmonary Angiography Image Simulation using Cycle Generative Adversarial Network from Chest CT imaging in Pulmonary Embolism Patients

Chia-Hung Yang[a], Yun-Chien Cheng[a, ‡], Chin Kuo[b,c, ‡]

[a] *Department of Mechanical Engineering, College of Engineering, National Yang Ming Chiao Tung University, Hsin-Chu, Taiwan*

[b] *Department of Oncology, National Cheng Kung University Hospital, College of Medicine, National Cheng Kung University, Tainan, Taiwan*

[c] *College of Artificial Intelligence, National Yang Ming Chiao Tung University, Hsin-Chu, Taiwan*

[‡]*The authors contributed equally to this work.*

*Corresponding author: yccheng@nycu.edu.tw , tiffa663@gmail.com

# Abstract

The purpose of this research is to develop a system that generates simulated computed tomography pulmonary angiography (CTPA) images clinically for pulmonary embolism diagnoses. Nowadays, CTPA images are the gold standard computerized detection method to determine and identify the symptoms of pulmonary embolism (PE), although performing CTPA is harmful for patients and also expensive. Therefore, we aim to detect possible PE patients through CT images. The system will simulate CTPA images with deep learning models for the identification of PE patients' symptoms, providing physicians with another reference for determining PE patients. In this study, the simulated CTPA image generation system uses a generative antagonistic network to enhance the features of pulmonary vessels in the CT images to strengthen the reference value of the images and provide a basis for hospitals to judge PE patients. We used the CT images of 22 patients from National Cheng Kung University Hospital and the corresponding CTPA images as the training data for the task of simulating CTPA images and generated them using two sets of generative countermeasure networks.
This study is expected to propose a new approach to the clinical diagnosis of pulmonary embolism, in which a deep learning network is used to assist in the complex screening process and to review the generated simulated CTPA images, allowing physicians to assess whether a patient needs to undergo detailed testing for CTPA, improving the speed of detection of pulmonary embolism and significantly reducing the number of undetected patients.

# Introduction

Pulmonary embolism (PE) is clinically difficult to diagnose because of its multifaceted symptoms. In addition, pulmonary embolism has multiple triggers and is one of the leading causes of vascular death, performing rapid detection and treatment can significantly reduce the risk of death in hospitalized patients. The current method of clinical detection of pulmonary embolism is by computerized tomography pulmonary angiography (CTPA), which is a CT scan that takes pictures of the blood vessels from the lungs to look for symptoms of pulmonary embolism. During the CTPA procedure, contrast is injected into the veins and then into pulmonary arteries. The contrast makes lung vessel appear bright and white on the scanned image to find out any blockages or blood clots which is dark in the image. However, the cost of taking a CTPA is higher and the risk of damaging the patients is increased. Therefore, generating simulated CTPA images from CT images of the lungs during clinical examination can increase the chance of detecting pulmonary embolism during the initial examination, and medical resources can be used more efficiently without having to take CTPA images of every patient to screen for PE. [1][2][3][4][5]

In our study, we first attempted to simulate CTPA images using interpolation to improve resolution and contrast adjustment, but the images generated by interpolation were blurred and could not improve recognition rate, and the images obtained by contrast adjustment were fragmented and could not be interpreted. In recent years, machine learning has become popular, and image generation has become a new research direction in the field of image processing[6], in addition to image classification and semantic segmentation using convolutional neural networks[7]. In medical imaging, Karim's team at Stuttgart University in Germany proposed MedGAN [8], a Generative Adversarial Network (GAN) network[9] applied to the medical imaging environment. They applied GAN to three different imaging environments, namely, conversion of positron tomography (PET) images to CT images, removal of image blurring during magnetic resonance (MR) photography, and noise reduction of PET images. In addition, Tien, HJ et al. published in 2021 the optimization of Cone-beam CT images by means of GAN [10]. In taking Cone-beam CT, the interference from X-ray scattering, noise and artifacts make Cone-beam CT images less clear, so the Cycle-Deblur GAN model was proposed to generate an image with more complete structural details and higher accuracy of CT values. This study successfully optimizes the details in CT images while preserving the original structural features. From the study of MedGAN and Cycle-Deblur GAN, we believe that generative adversarial networks can be used to generate valuable images in medical imaging and generating simulated CTPA images to assist in the diagnosis of PE should be a feasible study. From the aforementioned literature, we can use generative adversarial networks to generate images, change the style of the input images and highlight their specific features to achieve a certain degree of restoration in medical images. Therefore, we hope to simulate the CTPA images from the data of the initial PE test for doctors to use as a basis for clinical judgment. Among ECG, chest X-ray and CT images, CT images are the closest to CTPA images (Figure 1) and are relatively the most informative, so we choose to use CT images to generate simulated CTPA images.

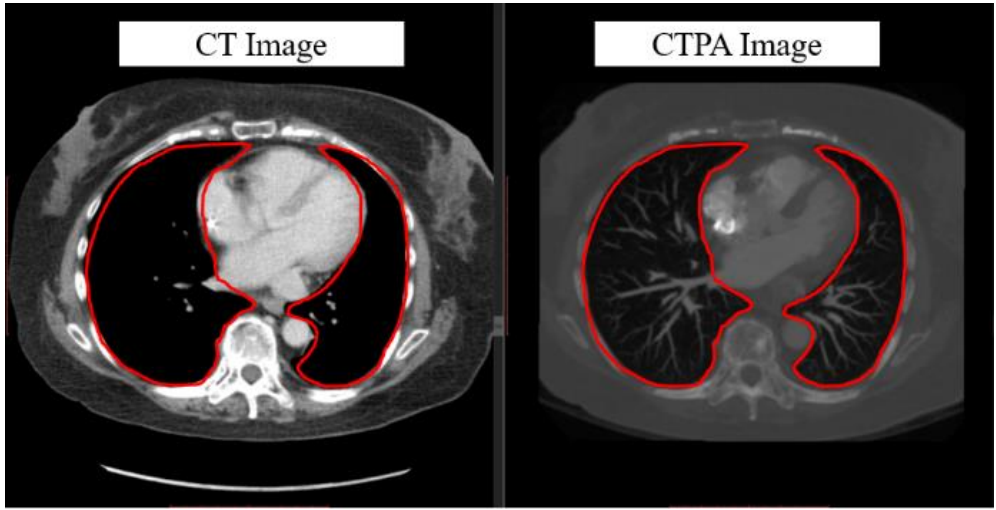

Figure 1. Comparison of CT images and CTPA images of the same cross-section

With our simulation system, the clinician can obtain a CTPA simulation immediately after the CT image is taken, and can schedule a CTPA test directly if a pulmonary embolism is suspected in the simulation, reducing the delay in the consultation process. If the simulated CTPA image does not show PE-related features, the patient can continue to follow the original procedure to rule out the possibility of PE, reducing the patient's risk of exposure to radiation in a single tomographic scan and reducing the harm to the patient from contrast injection. In addition, the simulated CTPA images can also be combined with many established PE computerized detection software (CAD) for PE diagnosis, further enhancing the accuracy of clinical diagnosis.(Figure 2.)

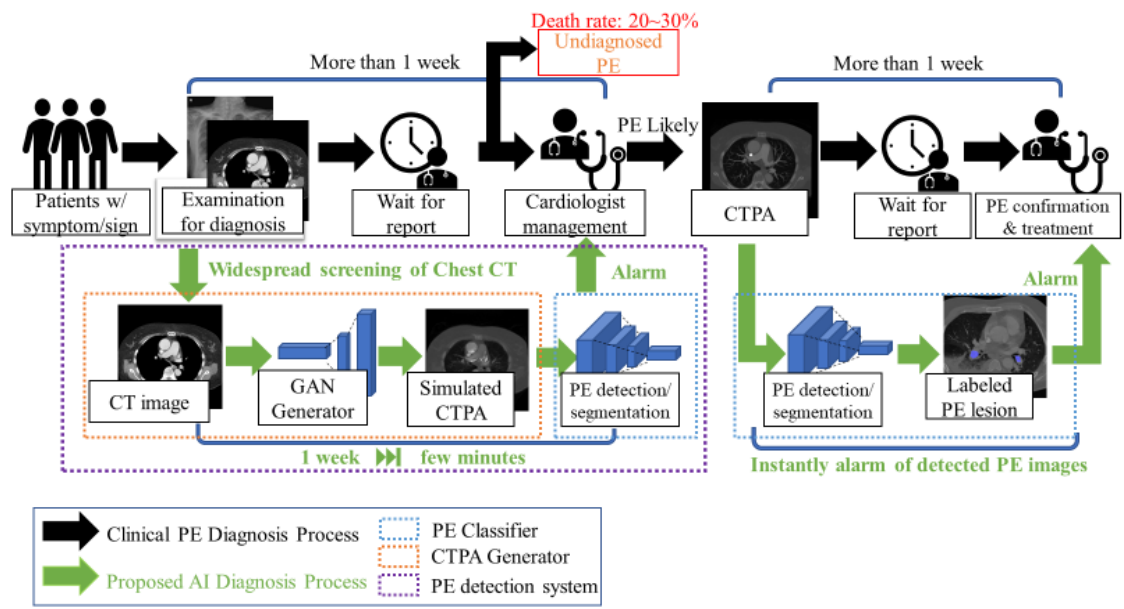

Fig. 2 Schematic diagram of the PE consultation process by generating simulated CTPA

# Materials and Methods

The first part of this study is the selection of the backbone architecture for generating the adversarial network model, and we use the Pix2pix architecture[11] to generate CTPA images from CT images and the CycleGAN architecture[12] to generate CTPA images. In the second part of the experiment, since we believe that the image dataset does not meet the target image alignment required by Pix2pix and other architectures, we choose CycleGAN, which does not require the matching of input and output images, to improve and discuss the details of the architecture.

**Image Sources**
The first part is a pre-experimental data set of 50 patients without pulmonary embolism from National Cheng Kung University Hospital as the backbone of the selection framework, and the second part is a data set of 22 patients with pulmonary embolism also from National Cheng Kung University Hospital. Both datasets were obtained from National Cheng Kung University Hospital and were reviewed by the Human and Behavioral Research Ethics Board of National Cheng Kung University School of Medicine (IRB No: B-ER-108-380). CT images and CTPA images were taken in these patients during diagnosis. Since there were some coordinate shifts and differences in slice detail between the CT and CTPA images, we used Velocity AI software to align the CT images with the coordinates of the CTPA images on the patient images without pulmonary embolism. However, we found that using image processing software to align the target images was equivalent to passing through another layer of simulation, and this simulation could not faithfully represent the original cross-sectional images.

**Data preprocessing**
In the first phase of the pre-experiment, we processed the images by dividing the DICOM image files into CT and CTPA as the input and output of the generation counterpart network, and converted the images into JPEG files for easy observation. We compressed the original image resolution from 512×512 to 256×256 to increase the learning capability of the model. In the second part of the dataset, we directly used DICOM file format images as input and selected only the image intervals with pulmonary embolism features as training input and target, in order to make the model focus on simulating the areas with pulmonary embolism in the CT images.

**First phase – pre-experiment**
**CTPA_Pix2Pix model**

The first model we designed is a modification of the Pix2pix architecture, using U-Net [13]as the generator network and two different discriminator networks for evaluation, a Pixel Discriminator for pixel-by-pixel inspection and a 3-layers Discriminator for distinguishing the characteristics of the output images, as shown in Figure 3. The patient's CT images and the corresponding CTPA images are matched and input to the network together, and the CT images enter the U-Net generator network to generate simulated CTPA images, and the discriminator network determines the difference between the simulated CTPA images and the input CT images, and sends the results back to the generation network for the next cycle of training.

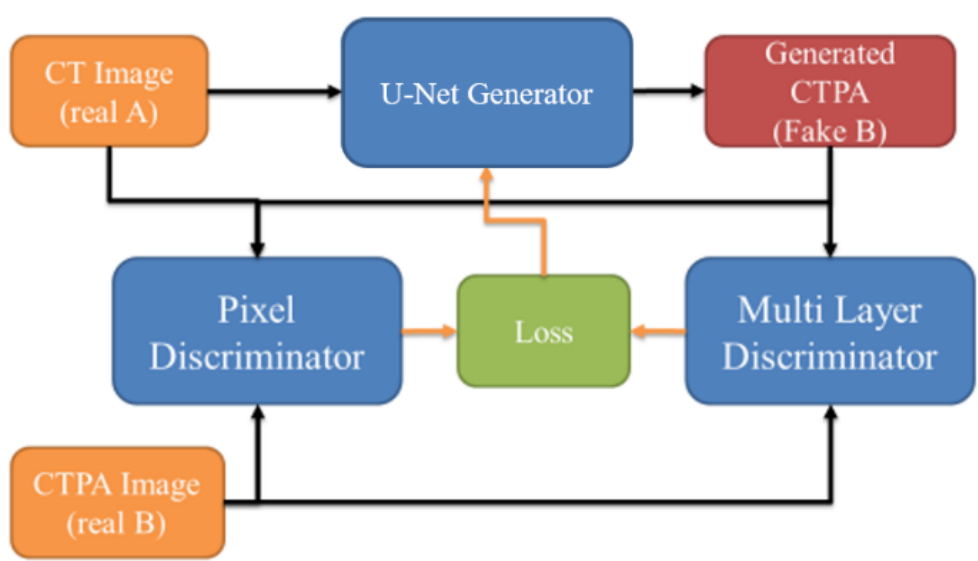

Figure 3. The CTPA_Pix2pix Model Framework

**CTPA_CycleGAN model**

Since the main difference between CTPA image and CT image is the characteristic of the contrast agent, and its characteristic can be regarded as a style conversion, we try to use the CycleGAN architecture to solve this problem.

We use two mirroring generator networks, as shown in Figure 4, generator network A is the network that converts CT images to CTPA images, and generator network B is the network that converts CTPA images back to CT images. First, the CT images are imported into the generator network A to form the simulated CTPA images, which are first evaluated by a discriminator network for their style conversion, and then imported into the generator network B to convert the simulated CT images back to the original input images for comparison. On the other hand, the CTPA images are input into the generator network B to generate the simulated CT images, and after judgment, they are sent to the generator network A to convert the simulated CTPA images and then compared with the original input CT images.

In view of the complexity of the image features that Cycle GAN may need to learn, this study uses the U-Net and ResNet [14] architectures to test the CycleGAN generator network architecture respectively.

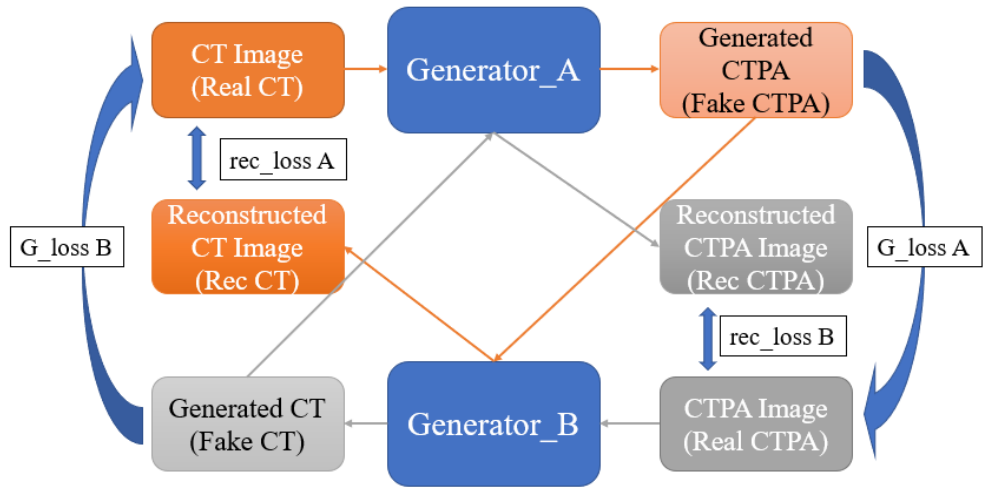

Fig. 4 The CTPA_Cycle GAN model architecture

**Second Phase**
**PE Classification CycleGAN**

Since we found in the first phase of experiments that aligning images is equivalent to generating a substandard simulation result as the target for generative adversarial network learning, we chose CycleGAN, which does not require image matching, as the backbone of our network.

In this phase of the experiment, we first tested the simulation of CTPA images using the original files in DICOM format as the input images. However, the DICOM format file covers a wide range of image values, which makes the model learning results not focus on the PE features of the pulmonary vasculature.

Therefore, we added a supervised neural network to the original CycleGAN to interpret the reconstructed CT images to see if they are images of pulmonary embolism, and to use this network to interpret these images as images of pulmonary embolism. In this way, our generative adversarial network can generate simulated CTPA images that are closer to those with pulmonary embolism features.

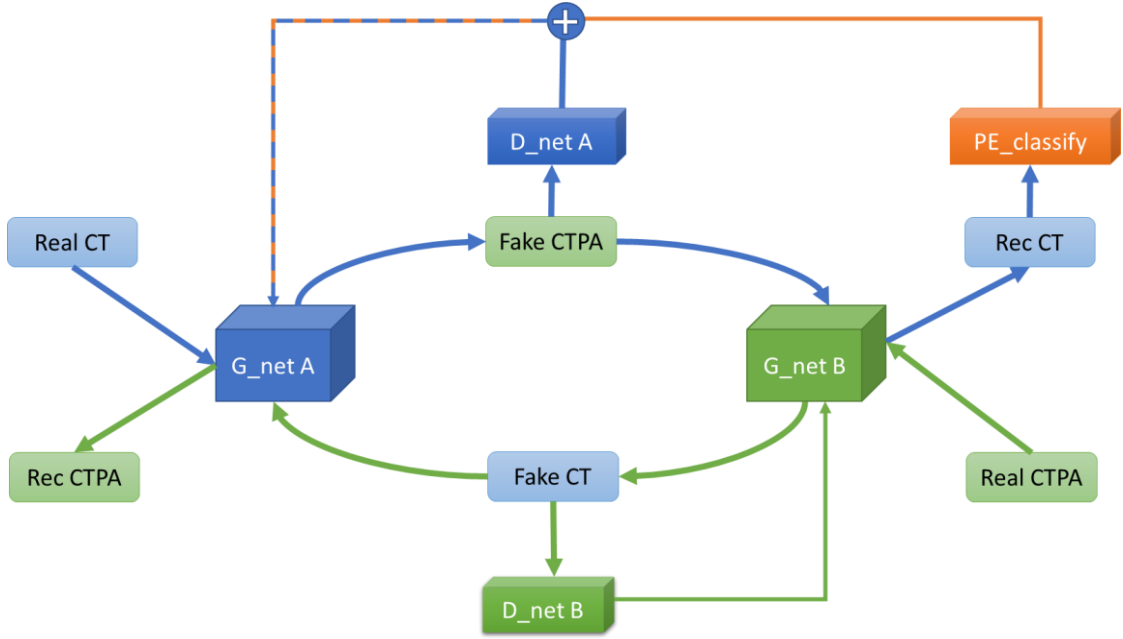

Figure 5.   PE Classification CycleGAN model architecture

Although the generated images are closer to the real CTPA images after adding the Pulmonary Embolism Recognition Network to the generative adversarial model for supervision, the CTPA image format is partially different from the CT image format, which is prone to structural deformation without pre-processing and alignment. We speculate that the main reason for this deformation is that our loss function uses Binary Cross Entropy loss (BCE loss) as the loss function of the generator and L1 loss as the loss function of the discriminator, and both loss functions are computed on a pixel-by-pixel basis . To solve this deformation problem, we try to use the structural similarity index (SSIM) as the discriminator loss function to increase the degree of network restoration to the structure and reduce the image caused by pixel deviation. We also try to use Mean absolute error (MSE) as the loss function of the generator to reduce the pixel deviation as well.

**Evaluation method**

Since the purpose of this experiment is to generate different types of medical images for conversion, there are no related studies, so this experiment will design a method to evaluate the results of our model generation. We hope that the model output images are as close to the original CTPA images as possible, so we want to observe the following objectives: image generation quality, structural similarity, and image similarity. Since the generated images are expected to have less noise, we use the peak signal-to-noise ratio (PSNR)[15] as a metric to evaluate the quality of the generated images. For PE, the structure of blood vessels is an important criterion, and we use the structural similarity index (SSIM) as the evaluation criterion to select the result that can best restore the structure. For image similarity, we will use the general image generation evaluation methods such as depth perception image similarity (LPIPS)[16] and Fréchet Inception Distance (FID) [17], and then calculate the mean absolute error (MAE) pixel by pixel to make a comprehensive evaluation.

Peak Signal-to-Noise Ratio (PSNR)
PSNR is a quantitative index used to evaluate the distortion of an image. the result of PSNR represents the ratio of the maximum possible signal power to the destructive noise power and is defined as follows:

$$\text{MSE} = \frac{1}{mn} \sum_{i=0}^{m-1} \sum_{j=0}^{n-1} [I(i,j) - K(i,j)]^2$$

$$\text{PSNR} = 10 \cdot \log_{10} \left( \frac{MAX_I^2}{MSE} \right)$$

If the size of both the generated image and the original image is m×n, and I is the original image and K is the generated image, the mean square error (MSE) of the two images can be obtained. In general, it is difficult to distinguish the difference between PSNR>30 and PSNR between 20 and 30, so that some difference can be felt. If the PSNR is between 10 and 20, it is obvious that there is noise, but the similarity of the two images can still be seen. If the PSNR is below 10, it is difficult for the human eye to determine the similarity of the images.

Structural Similarity Index (SSIM)
SSIM is similar to PSNR in that it is also an indicator of the quality of image production. However, unlike PSNR, SSIM places more emphasis on structural information []. In a natural image, there should be a strong correlation between adjacent pixels, and such a correlation can express the structural information in the image scene. Therefore, SSIM is defined as follows:

$$\text{SSIM}(x, y) = [l(x,y)]^\alpha [c(x,y)]^\beta [s(x,y)]^\gamma$$

$$l(x, y) = \frac{2\mu_x \mu_y + C_1}{\mu_x^2 + \mu_y^2 + C_1} \quad c(x, y) = \frac{2\sigma_x \sigma_y + C_2}{\sigma_x^2 + \sigma_y^2 + C_2} \quad s(x, y) = \frac{\sigma_{xy} + C_3}{\sigma_x \sigma_y + C_3}$$

In SSIM, three parameters are evaluated: luminance l(x,y), contrast c(x,y) and structure s(x,y), with C_1, C_2 and C_3 as constants. If two identical images are computed by

SSIM, it will get 1. Therefore, the closer the SSIM index is to 1, the higher the similarity of the two images.

Deep Image Perception Similarity (LPIPS)

For the human eye, it is easy to quickly evaluate the perceptual similarity between two images, but this evaluation process is not well quantified. The human eye does not evaluate the similarity of two images on a pixel-by-pixel basis. After the significant application of neural networks in recent years, Richard Zhang et al. found in 2018 that it is useful to extract feature maps of shapes through VGG networks as a basis for image judgment, and therefore proposed a new evaluation method, LPIPS, to systematically compare the deep features in different images [].

As shown in Fig. (), LPIPS is calculated by inputting two images to a VGG network, obtaining the vector map of the specific convolutional layer of the two images in the VGG network and calculating the remaining chordal distances, and then averaging these distances to obtain the LPIPS score.

Fréchet Inception Distance (FID)

FID is a common evaluation method for GAN image generation, which calculates the distance between the real image and the feature vector of the generated image as an indicator of the quality of the image generation []. The FID score uses the classification model of Inception v3 and takes the last pooling layer as the feature map for evaluation, and by calculating the mean and standard deviation between these image feature maps, the target image and the generated image are generated as a set of After calculating the mean and standard deviation between these image feature maps, a Gaussian distribution is generated for the target image and the generated image, and the distance between these two distributions is calculated by the Wasserstein-2 method. Therefore, the FID score should be 0.0 in the best case, which means that the distributions of the two sets of images are exactly the same.

## Results

**First Phase - Generating Contrastive Model Architectures for Comparison**

We first exported and compared the results generated by the unmodified model after 1500 epochs of training. We tested the CT images of the patients and their matching CTPA images with independent untrained images, and only the trained generation network was used as the network for image generation.

The results of the CTPA_Pix2pix model are shown in Figure 6. We can find that the images generated by the generator perform well in the reproduction of bones and organs, with little error from the real images, but are relatively blurred in the judgment of the PE block, except for the thicker arteries and veins, and the vessels of the lungs are very poorly imaged.

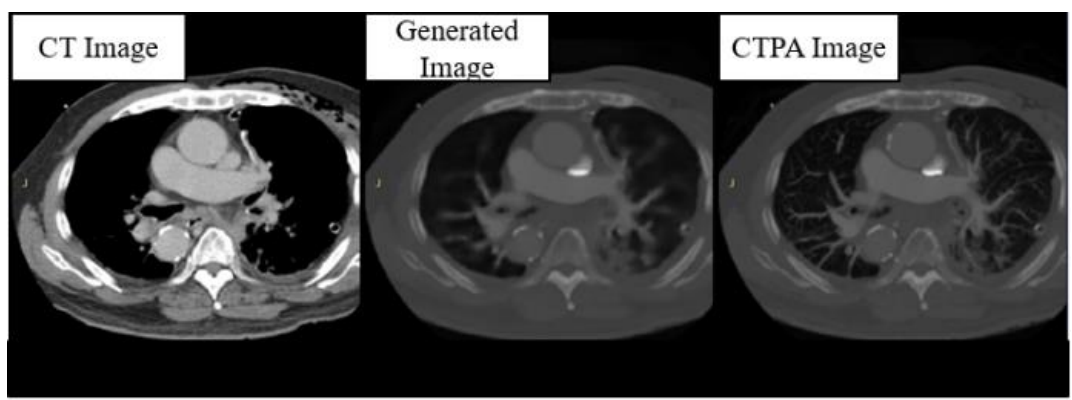

Fig. 6 Simulated images generated by CTPA_Pix2pix model

In the CTPA_CycleGAN model, as shown in Figure 7, we used U-Net and ResNet as the backbone of the network for training, and we can see that the performance on U-Net is not as good as expected, although the restoration performance on bones and organ tissues is excellent as in the CTPA_Pix2pix model, but the microvascular part of the lungs The generation of the lung microvasculature was different. In the CTPA_CycleGAN model (U-Net), the amount of vessel restoration was improved, but the main pulmonary artery was not generated, making the vascular trend different from reality and making it difficult to identify the symptoms of PE.

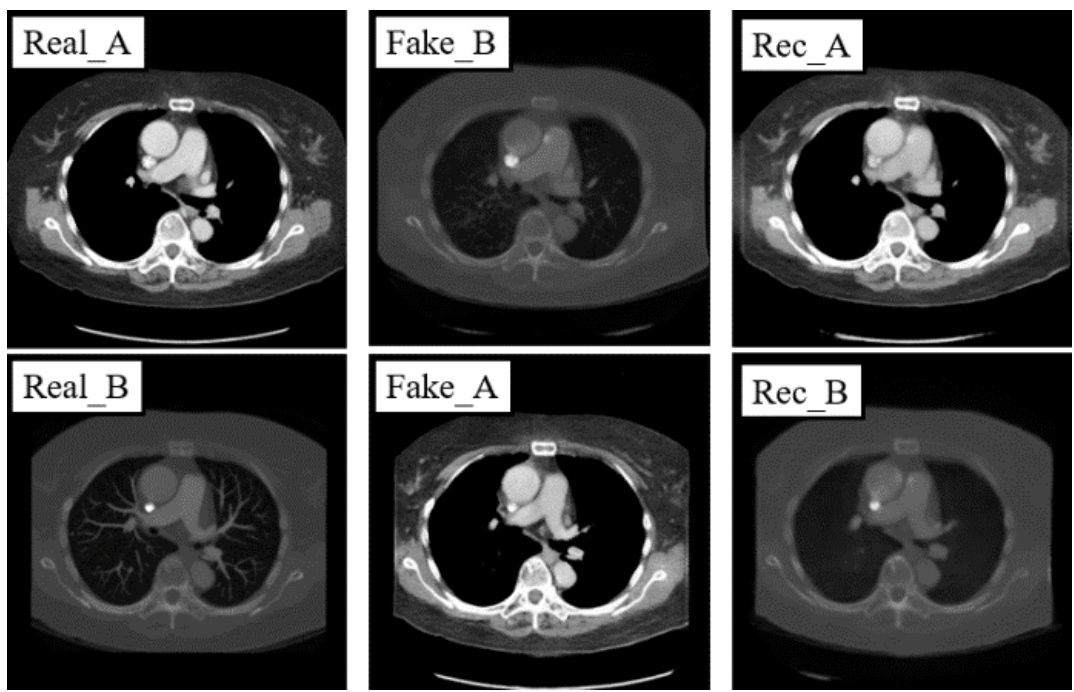

Fig. 7 Simulated images generated by CTPA_CycleGAN model (U-Net)

In the generation results of CTPA_CycleGAN model (ResNet), as shown in Figure 8, we can find that the simulated CTPA image Fake_B generated by generator A has a high degree of similarity to the real CTPA image Real_B, and the vascular imaging is also the best among all models. A point that is worth discussing is the Rec_B image, which is a simulated CTPA image generated by the CTPA image through generator network B and then generated by generator network A. Its similarity to Real_B is even higher than that of Fake_B, and there is almost no difference with the original image.

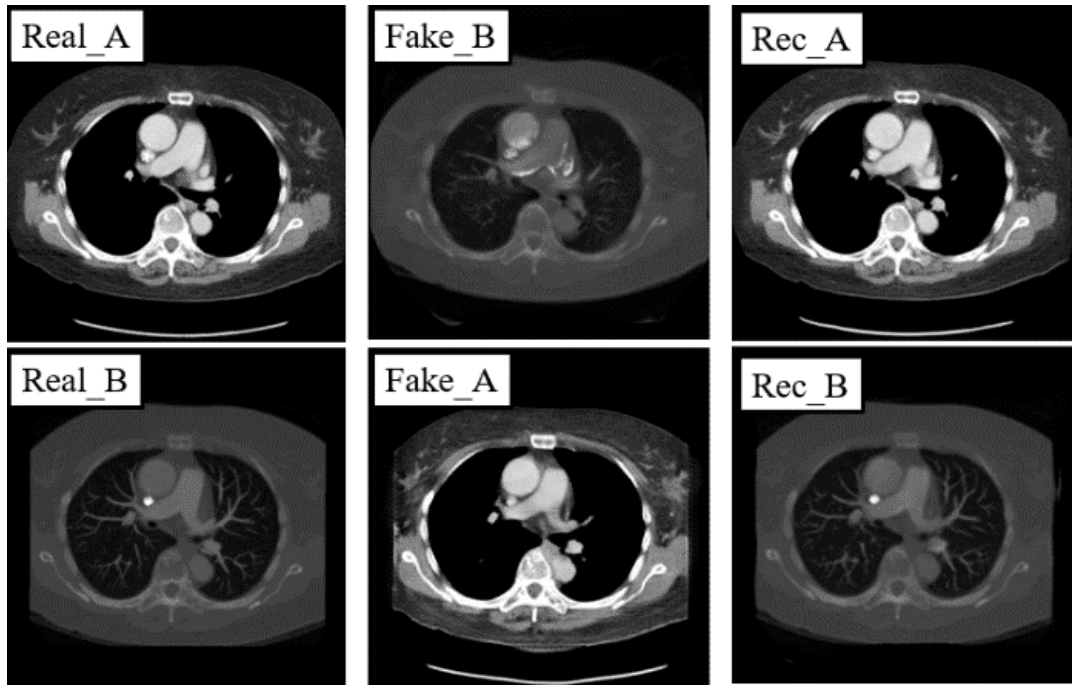

Fig. 8 Simulated images generated by CTPA_CycleGAN model (ResNet)

Second Phase - Results of CycleGAN model supervised with pulmonary embolism classifier

In the comparison of generative adversarial model architectures, we can find that CycleGAN performs the best in the task of converting CT images to CTPA images, and in clinical applications, CT images without imaging software simulation will not have corresponding CTPA images to be used as the generated target. Therefore, the DC GAN and Pix2pix architectures, which must use control sets, cannot be used as the generative adversarial network architecture for this experiment. For these two reasons, we decided to use CycleGAN as the backbone network architecture for CT generation simulation of CTPA. In this section, we will investigate the differences in the output results of different adaptations and modifications of CycleGAN and the reasons for them. Since most of the models we evaluate are pixel-based, we first fine-tune the coordinates and dimensions of the output images to align them with the original images when evaluating the generated images.

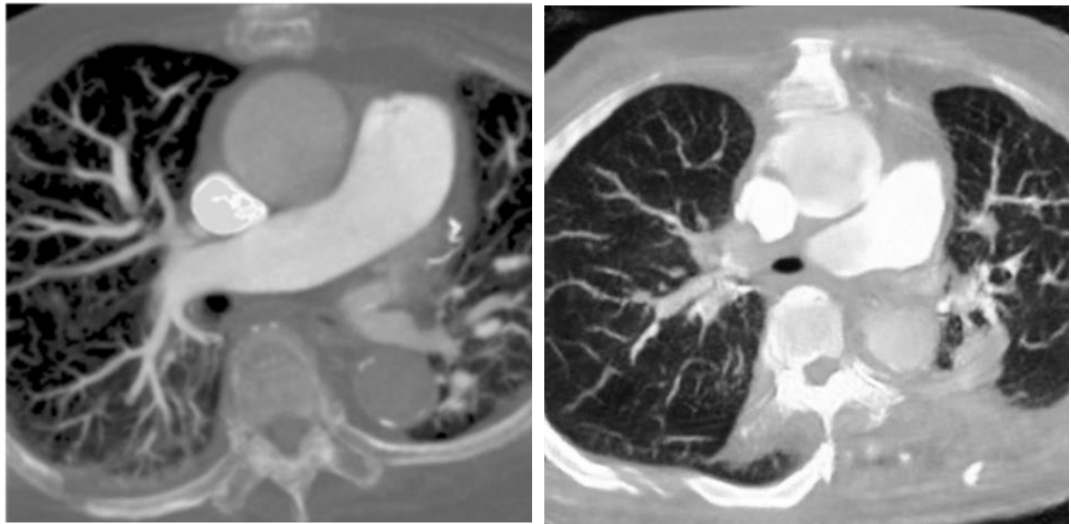

Fig. 9 Comparison of the generated results. The right graph is the simulated cross section generated by the proposed generator and the left is the original cross section.

In the generated results in Figure 9, we can see that the generated model reproduces the vascular vein as much as possible, but there are still many differences that are visible to the naked eye, and we will discuss how the model is adjusted later.

Table1. Comparison of the generated results

|       | Simulated CTPA |
|-------|----------------|
| PSNR  | 11.24          |
| SSIM  | 0.324          |
| MAE   | 102.13         |
| LPIPS | 0.439          |
| FID   | 223.68         |

# Discussion

We compared CycleGAN networks with different generators and found that the performance of the generated results on ResNet was significantly better than that on U-Net, and the reconstruction loss and generator loss were both better than those on U-Net. In terms of reconstruction loss, the deeper ResNet50 performs better than ResNet9. The loss function represents the ability to restore the graphs, and the source of comparison is the original graph and the graphs restored by two layers of generators. It can be assumed that this generative feature is more diverse and therefore more complex networks can be considered as generator networks in the future. However, the more complex network may be relatively difficult to train in an adversarial way, so we plan to pre-train it by classifier first, and then place it into the generative adversarial network for optimization.

First, we input the original DICOM images for the first training, and found that the CT values of each pixel in the DICOM file were too widely distributed, which caused the model output to fail to focus on the PE features. Since our goal is to detect PE, we try to pass the input images through a HU filter first, and keep the relevant CT values of lungs and blood vessels into the model to enhance the model's ability to restore PE features. In addition, since CycleGAN is an unsupervised adversarial generative neural network, we believe that if we can add a pre-trained classification network for PE features in its discriminator, it can enhance its ability to restore PE features. In Fig. 10, we can see that with the addition of the classification network, CycleGAN can focus more on the vascular features and lung regions for reproduction, and the overall image similarity is greatly improved. In each of the image evaluation indexes (Table 2), we can see that the generated network with the addition of the classification model not only has less noise and higher structural similarity, but also has improved the overall image style similarity.

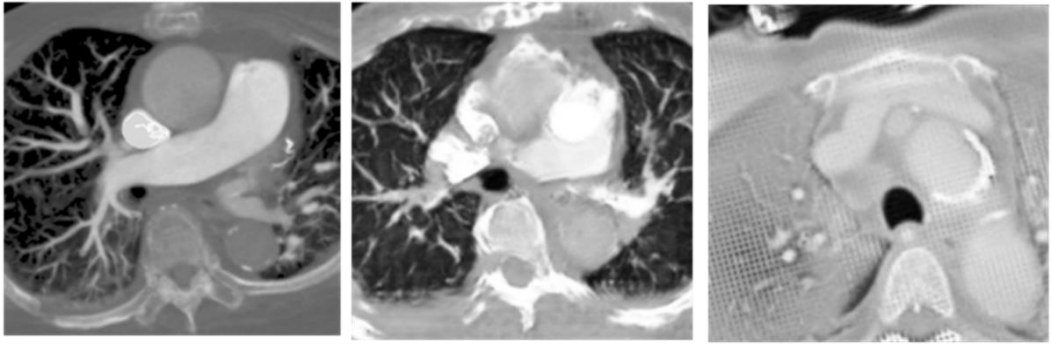

Fig. 10 Comparison of the output image with categorization network. The left is the original image, the middle is the result of adding classification network to the analog generation network, and the right is the result of the analog generation network without classification network.

Table 2 Comparison of the output image with categorization network

|       | Classification model | Without classification model |
|-------|----------------------|------------------------------|
| PSNR  | **11.227**           | 7.79                         |
| SSIM  | **0.325**            | 0.207                        |
| MAE   | **96.9**             | 114.72                       |
| LPIPS | **0.494**            | 0.526                        |

In the second stage, we compare the weights of the loss returned to the generator by the classification network and CycleGAN. We find that if the returning weight of the classification network is too high, the generated image will easily resemble the feature map of the classification network due to the lack of the adjudicator to monitor the quality of the generated image. Therefore, we multiply the categorical network return value by a weight to reduce its image quality, but still retain the ability to monitor the PE features. From Fig. 11, we can find that a weight between 0.1 and 0.3 results in the best image restoration ability, and from the values in Table 3, we can see that as the weight decreases, the structural similarity is lost due to the decrease in the requirement for PE features, and the rest of the values do not differ significantly.

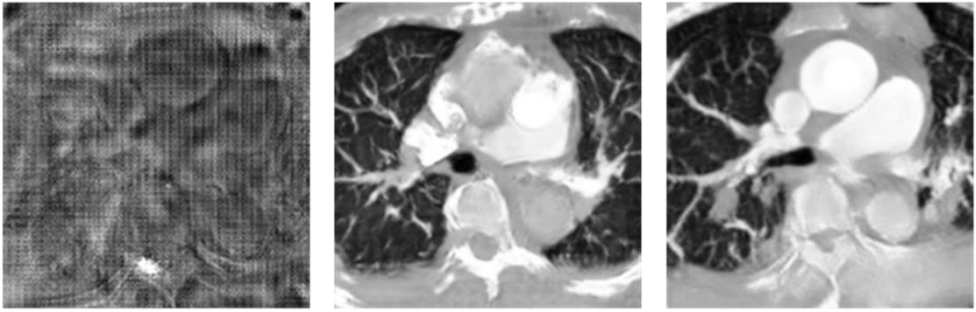

Fig. 11 Comparison of classification network and discriminator weights on the output image. The weights from left to right are 1, 0.3, and 0.1 respectively.

Table 3 Comparison of classification network and discriminator weights on the output image.

|       | Ratio = 1 | Ratio = 0.3 | Ratio = 0.1 |
| ----- | --------- | ----------- | ----------- |
| PSNR  | 11.19     | 11.227      | **11.23**   |
| SSIM  | 0.116     | **0.325**   | 0.304       |
| MAE   | 122.72    | 96.9        | **96.38**   |
| LPIPS | 0.540     | 0.494       | **0.428**   |

In adversarial generative networks, the discriminator is also an important basis. If the discriminator is too powerful, the generator may not learn easily. Therefore, we try to balance the learning ability of generators and discriminators, but due to the hardware limitation, our experimental platform cannot handle such a huge computation if the generators are deeper, so we fix the generator as ResNet34 and adjust the number of layers of discriminators to achieve a balance in training. In this experiment, we tested the discriminators of three-, four-, and six-layer NNs, and we found that the discriminator with six layers was too fast to learn the convergence of the generative network, so we only compared the results of the three- and four-layer discriminators. As shown in Figure 12 and Table 4, we believe that the combination of the three-layer discriminator and ResNet34 is the most suitable for the CT to CTPA conversion task.

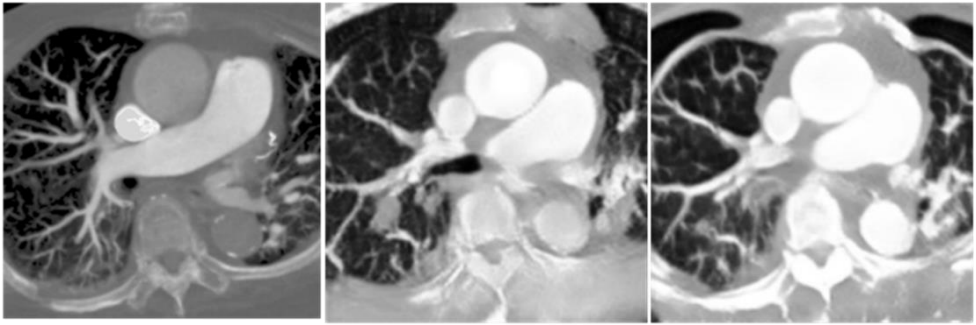

Fig. 12 Comparison of the output image with different depths of the discriminators. The left image is the original image, the middle image is the result of training the discriminator with three layers of depth, and the right image is the result of training the discriminator with four layers of depth.

Table 4 Comparison of different depths of the discriminators on the output images

| Target CTPA | 3-layers | 4-layers |
|---|---|---|
| PSNR | 11.23 | **11.25** |
| SSIM | **0.304** | 0.276 |
| MAE | **96.38** | 97.31 |
| LPIPS | **0.428** | 0.459 |

After adjusting the main model structure, we found out that the generated model tended to enlarge the simulated image due to the size difference between the CT image and the CTPA image in the original format. This action caused distortion of the image structure. Therefore, we adjusted the loss function calculation of the generator and the discriminator respectively. In the discriminator, we replaced the L1 loss by SSIM to increase the discriminator's requirement for the image structure, while in the generator, we tried to use MSE to complement the pixel-to-pixel similarity. From Fig. 13 and Table 5, we can find that the use of SSIM as the discriminator alone will cause the degradation of the generated image quality.

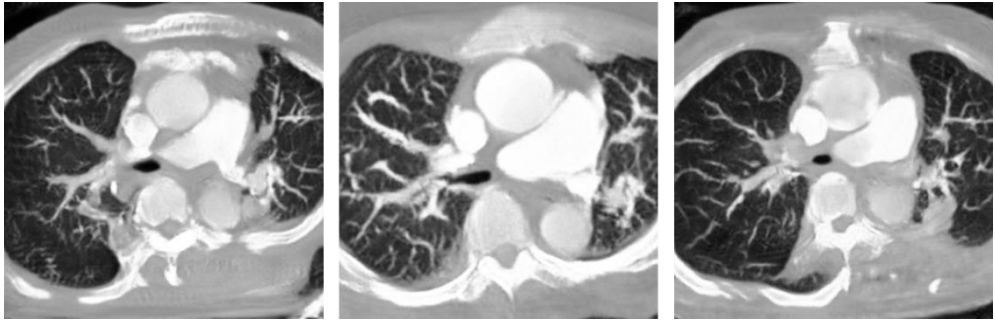
Fig. 13 Comparison of the loss function on the output image. The left figure shows the result of using BCE loss with L1 loss, the middle figure shows the result of BCE loss with SSIM, and the right figure shows the result of MSE loss with SSIM.

Table 5 Comparison of loss function for output image

|      | BCE+L1 | BCE+SSIM | MSE+SSIM |
|------|--------|----------|----------|
| PSNR | 11.24  | **11.25**    | 11.24    |
| SSIM | 0.309  | 0.171    | **0.324**    |
| MAE  | **96.84**  | 106.76   | 102.13   |
| LPIPS| 0.479  | 0.510    | **0.439**    |

Finally, we compare the results of PE classification network supervising different sources on CycleGAN. Since our classification network wants to preserve the PE features in the CT images, we can supervise two images: the reconstructed CT images generated from the simulated CTPA images and then transformed by CycleGAN (Rec_CT) and the simulated CT images generated from the CTPA images (Fake_CT). From Figure 14 and Table 6, we can find that the images generated by Rec_CT supervised by PE classification network are more similar to the original images, while the output results supervised by Fake_CT are more likely to produce structures that are not present in the original images. We speculate that the source of the PE features supervised on Fake_CT is the CTPA image which is not the original CT image, so this PE classification network will prompt the generator to generate the vessels that may generate PE features by generalizing from the CTPA image, instead of emphasizing the PE features from the original image.

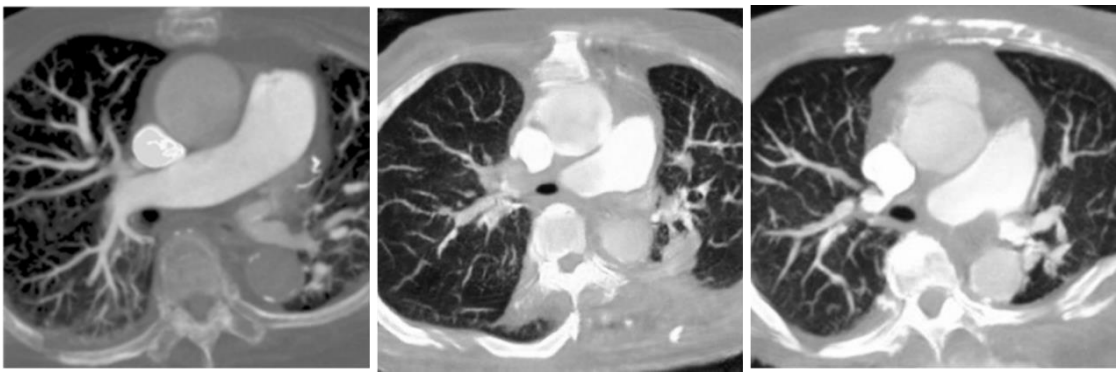

Figure 14 Comparison of categorized network monitoring targets to output images

Table 6 Comparison of categorized network monitoring targets to output images

| Target CTPA | On Rec_CT | On Fake_CT |
|---|---|---|
| PSNR | **11.24** | 11.23 |
| SSIM | 0.324 | **0.336** |
| MAE | **102.13** | 107.44 |
| LPIPS | **0.439** | 0.495 |
| FID | **223.68** | 250.02 |

Observing the above experimental data, we can see that the parameter PSNR for judging image quality is between 10 and 20, and this value is not yet judged to be similar in image quality. The reasons for this may be the following. First, the PSNR index is originally designed to determine the image compression quality, and the images are all neatly aligned, and the difference between the PSNR and the original image is a small difference in the compressed image algorithm. However, when applied to the image generation field, especially in the style-shifting CycleGAN framework, it is easy to cause significant PSNR degradation due to small image misalignment. In order to solve this problem, we try to minimize the parameter discrepancy caused by this bias by manual correction. Secondly, the result of image conversion is not from the same image as the compressed image, and there are some resolution and size differences between our CTPA images and CT images, which makes the PSNR results not always better. In addition, our SSIM results also have similar problems to PSNR, and we have manually corrected them to reduce the error, but there is still the problem of different image standards. In summary, we believe that the LPIPS and FID scores of the deep learning framework are more suitable as the reference standard for the evaluation of image generation tasks.

In summary, we believe that the best method to simulate CTPA images with PE features is to enhance the supervised CycleGAN architecture by a PE classification network with 0.3 times of weight on the reconstructed CT images, and choose ResNet50 for the generator and a three-layer convolutional neural network for the discriminator. The loss function is chosen as MSE as the generation loss and SSIM as the discriminant loss, and the initial input images should be filtered by HU filter. For the final model evaluation, LPIPS and FID score are more accurate criteria to evaluate the generated model.

In our current study, we can see that the CT images can be generated by generating a counter network to produce CTPA images, and most of the bones, organs and body structures can be reproduced completely. generation ability is not as expected. In CTPA_CycleGAN, the model with ResNet as the skeleton can enhance the generation of microvessels, but the microvessels are not derived from the thicker major vessels but directly from the cavities when generating microvessels. After adding the PE classifier, our proposed PE classification CycleGAN can generate simulated CTPA images from CT images more accurately. The current image generation results can still be improved in two directions. First, the number and diversity of training data can be increased by adding datasets from different hospitals and different regions, so that the model can learn more different pulmonary embolism features and lung regional structural features to increase the ability of the model to restore images. In addition, the output image results still need to be tested clinically in the hospital, and the resulting images will be interpreted by the physicians to observe whether the symptoms of pulmonary embolism can be successfully detected, and then the model structure will be further adjusted to generate simulated images with more reference value for clinical judgment.